\documentclass[singlecolumn,superscriptaddress,aps,floatfix,11pt]{revtex4-2}
\usepackage{hyperref}
\usepackage{amsmath}
\usepackage{amssymb}
\usepackage{graphicx}
\usepackage{mathrsfs}
\usepackage{braket}
\usepackage{bm,bbm}
\usepackage{color}
\usepackage{multirow}
\usepackage{textcomp}

\newcommand{\br}{\bm{r}}

\begin{document}
\title{Dynamical Superfluid and Bose-Insulator Phases in Quantized Polariton Lattices}
\author{Sanjib Ghosh}
\email{sanjibghosh@cuhk.edu.cn}
\affiliation{School of Science and Engineering, The Chinese University of Hong Kong (Shenzhen), Longgang, Shenzhen, Guangdong, 518172, P.R. China}
\affiliation{Beijing Academy of Quantum Information Sciences, Beijing, 100193, P. R. China}
\date{\today}

\begin{abstract} 
 We demonstrate that Hilbert-space quantization in polariton lattices—manifested as multiple quantized energy levels in strongly confined sites—provides an unconventional route to realizing and manipulating different quantum phases. We show that nonlinear interactions transfer population into excited on-site quantum levels, which acts as an intrinsic dynamical channel controlling quantum coherence across the lattice. While weak nonlinearity confines polaritons to the lowest mode, yielding a robust superfluid phase with broken U(1) symmetry, strong nonlinearity induces phase diffusion through inter-level mixing. This dynamically generated fluctuations suppress global phase coherence and drives the system into a dynamical Bose-insulating phase. The changes between these phases occurs either as a nonequilibrium phase transition or a sharp crossover.
\end{abstract}

\maketitle

\textbf{Introduction:-- } 
Controlling quantum states of matter between ordered and disordered phases is a ubiquitous theme in condensed-matter physics, where symmetry-protected macroscopic order can be tuned by external fields or drives~\cite{Domb2000}. Traditionally, such phenomena are driven by thermal fluctuations, allowing systems to explore different configurations. However, thermal processes are often slow and difficult to implement in compact or integrated platforms. In contrast, non-thermal mechanisms—including quantum fluctuations~\cite{Sachdev1999}, external driving, and optical control—provide more practical routes to manipulate quantum states and realize novel phases of matter.

A paradigmatic example is the superfluid–Mott insulator transition, where interactions can transform a phase-coherent superfluid into an incompressible insulating state with precisely integer occupation at each lattice site~\cite{Bloch2022}. Similar phenomena occur in superconductors, where tuning fields or impurities can suppress global phase coherence and induce a Bose-insulating phase—without breaking Cooper pairs, unlike the conventional superconductor–metal transition~\cite{Fisher1989,Wang_2024}. These examples highlight a unifying principle: the emergence and loss of order is governed by the interplay of coherence, interactions, and the structure of the Hilbert space~\cite{Sondhi1997,Vojta2003}.

Exciton–polaritons, composite bosons formed through strong coupling between photons and excitons~\cite{Carusotto2013,Kavokin2017}, provide a promising platform for exploring different physical phenomena~\cite{Dagvadorj2015,Ghosh2022}. Like Cooper pairs, polaritons acquire phase coherence through condensation~\cite{Deng2010,Byrnes2014}, making polariton lattices natural candidates for realizing quantum coherent phases and their transitions~\cite{Brune2025}. However, a key challenge remains: such quantum phenomena typically require well-defined Fock-space quantization~\cite{Greentree2006,Liew2023}. Although polaritons are Bosonic quasi-particles, their finite lifetime—and the associated linewidth—prevents particle-number quantization from surviving long enough to drive between different quantum states~\cite{Byrnes2010,Na2010}.

Recent advances in polaritonic technologies have significantly expanded the accessible parameter space, enabling precise optical and electrical control~\cite{SongKok2025}, robust topological~\cite{Karzig2015,Bardyn2015,Klembt2018,Wu2023} and non-Hermitian states~\cite{Hu2024,Krol2025,Bao2025,Xu2025,Shi2025}, high-quality coherent dynamics~\cite{Jia2025}, and innovative applications~\cite{Matuszewski2024,Barrat2024,Opala2025,Wouter2025,Sedov2025}. Polaritonic lattices are usually implemented using periodic potentials, where each site forms a local well with tunable inter-site hopping governed by the site separation~\cite{Jean2017,Su2020}. Importantly, strongly confined sites naturally host multiple quantized energy levels, although these internal modes are often neglected because polariton condensation is typically desired in a single dominant mode~\cite{Amo2016,Su2020,Masharin2025}.

In this work, we show that these internal quantized levels constitute a resource for engineering dynamical quantum phases in polariton systems. We demonstrate that Hilbert-space quantization—through the presence of multiple on-site modes—opens new quantum channels that can either stabilize global phase coherence or induce strong phase fluctuations. In the weakly interacting regime, polaritons predominantly occupy the ground state, whereas higher states dynamically stabilize the superfluid phase. In contrast, strong nonlinear interactions activate substantial mode mixing, which drives phase diffusion and ultimately destroys long-range coherence, yielding a dynamical Bose-insulating phase. Crucially, we find that lattices lacking such quantized levels fail to exhibit this transition, underscoring the essential role of Hilbert-space structure.

Our results uncover a dynamical mechanism affecting the long-range order in polariton lattices, showing that internal mode quantization—rather than particle-number quantization—can influence the stability of phase coherence by modifying the U(1) gauge symmetry at long wavelengths. This mechanism gives rise to a dynamical phase transition, or a sharp crossover. This establishes exciton–polariton condensates as a versatile platform for realizing and controlling collective phases through intrinsic Hilbert-space engineering.

\begin{figure}
\includegraphics[width= 0.9\linewidth]{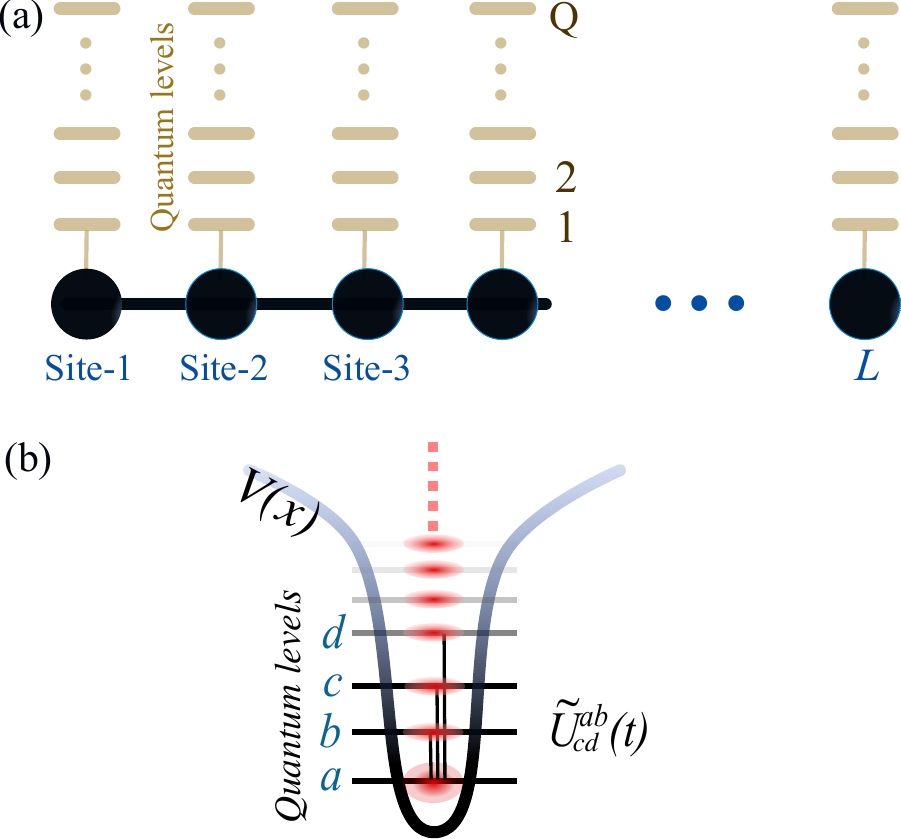}
\caption{A lattice with internal quantized energy levels due local potential minima. (a) A lattice with $L$ sites and $Q$ quantum levels per site. (b) A local potential minimum holding multiple quantized energy levels allowing mode mixing through the interaction coefficients $\tilde{U}^{ab}_{cd}(t)$ between the levels $a,b,c,$ and $d$.}
\label{fig:LatticeScheme}
\end{figure}

\textbf{The Model:-- } We begin from the driven-dissipative nonlinear Schr\"odinger equation~\cite{Wouters2007}
\begin{eqnarray}
    \label{eq:PDE}
  i\hbar\dot{\psi}(\mathbf r,t)
  = \left[-\frac{\hbar^2}{2m}\nabla^2 + V(\mathbf r)
    + \frac{i}{2}(P - \gamma)\right]\psi(\mathbf r,t) \notag \\
    + g\,|\psi(\mathbf r,t)|^2\psi(\mathbf r,t),
\end{eqnarray}
where $V(\mathbf r)$ denotes a periodic potential, $\gamma$ is the loss rate, $P$ represents the incoherent pump, and $g$ is the Kerr-type nonlinear interaction coefficient.

We consider a configuration as shown in Fig.~\ref{fig:LatticeScheme}(a), where $\{w_{i\alpha}(\br)\}$ is set of orthonormal, spatially localized orbitals centered at lattice sites $i=1,\dots,L$ and representing quantized energy levels $\alpha=1,\dots,Q$. The local quantized modes are taken as the eigenstates of an Hamiltonian $H_\text{loc} = -\frac{\hbar^2}{2m}\nabla^2 + V([\br])$ where $[\br]$ defines the region within a unit cell, satisfying $ H_\text{loc}| w_{i\alpha} \rangle
= E_{\alpha}\, \ket{w_{i\alpha}}$
with corresponding eigenenergies $E_{\alpha}$ and real orthonormal mode functions $\ket{w_{i\alpha}}$. We expand the condensate field in this Wannier-like basis as $\psi(\mathbf r,t) = \sum_{i=1}^{L}\sum_{\alpha=1}^{M}\psi_{i\alpha}(t)\,w_{i\alpha}(\mathbf r)$, where $\psi_{i\alpha}(t)$ are the complex mode amplitudes associated with site $i$ and level $\alpha$.

To describe the dynamics of the effective lattice, we project Eq.~\eqref{eq:PDE} onto the local orbitals $\bra{w_{j\beta}}$, yielding a discrete nonlinear Schrödinger equation for the modal amplitudes. Furthermore, to remove the rapid phases associated with $E_\alpha$, we introduce the rotating-frame transformation $\psi_{j\alpha}(t)=a_{j\alpha}(t)\,e^{-iE_\alpha t/\hbar}$. Finally, the discretized form of Eq.~\ref{eq:PDE} describing a lattice is given by,
\begin{equation}
i\hbar \dot{a}_{i,\alpha}
= -K \sum_{\langle ij \rangle } a_{j,\alpha}
+ g\sum_{\beta\mu\nu}
\tilde{U}_{\alpha\beta}^{\mu\nu}(t)\,
a_{i,\beta}^* a_{i,\mu} a_{i,\nu},
\label{eq:final_discrete_SE}
\end{equation}
where an external pump compensates the losses, so that a stable population is maintained (a more general case is discussed later), and the two terms respectively describe inter-site tunneling with an amplitude $K$, and nonlinear onsite interactions with $\tilde{U}_{\alpha\beta}^{\mu\nu}(t)
= U_{\alpha\beta}^{\mu\nu}\,
e^{i(E_\alpha+E_\beta-E_\mu-E_\nu)t/\hbar}$. The interaction tensor $U_{\alpha\beta}^{\mu\nu}$ is given by,
\begin{equation}
U_{\beta\alpha}^{\mu\nu} = \!\int\! d\mathbf r\,
w^*_{i\beta}(\mathbf r)\,
w^*_{i\alpha}(\mathbf r)\,
w_{i\mu}(\mathbf r)\,
w_{i\nu}(\mathbf r) 
\end{equation}
which is schematically shown in Fig.~\ref{fig:LatticeScheme}(b). In general, Eq.~\eqref{eq:final_discrete_SE} therefore provides a unified multimode framework that captures different phases of polaritons due to the competition between tunneling ($K$), and onsite nonlinearity ($g$).

\begin{figure}[h]
\includegraphics[width=1\linewidth]{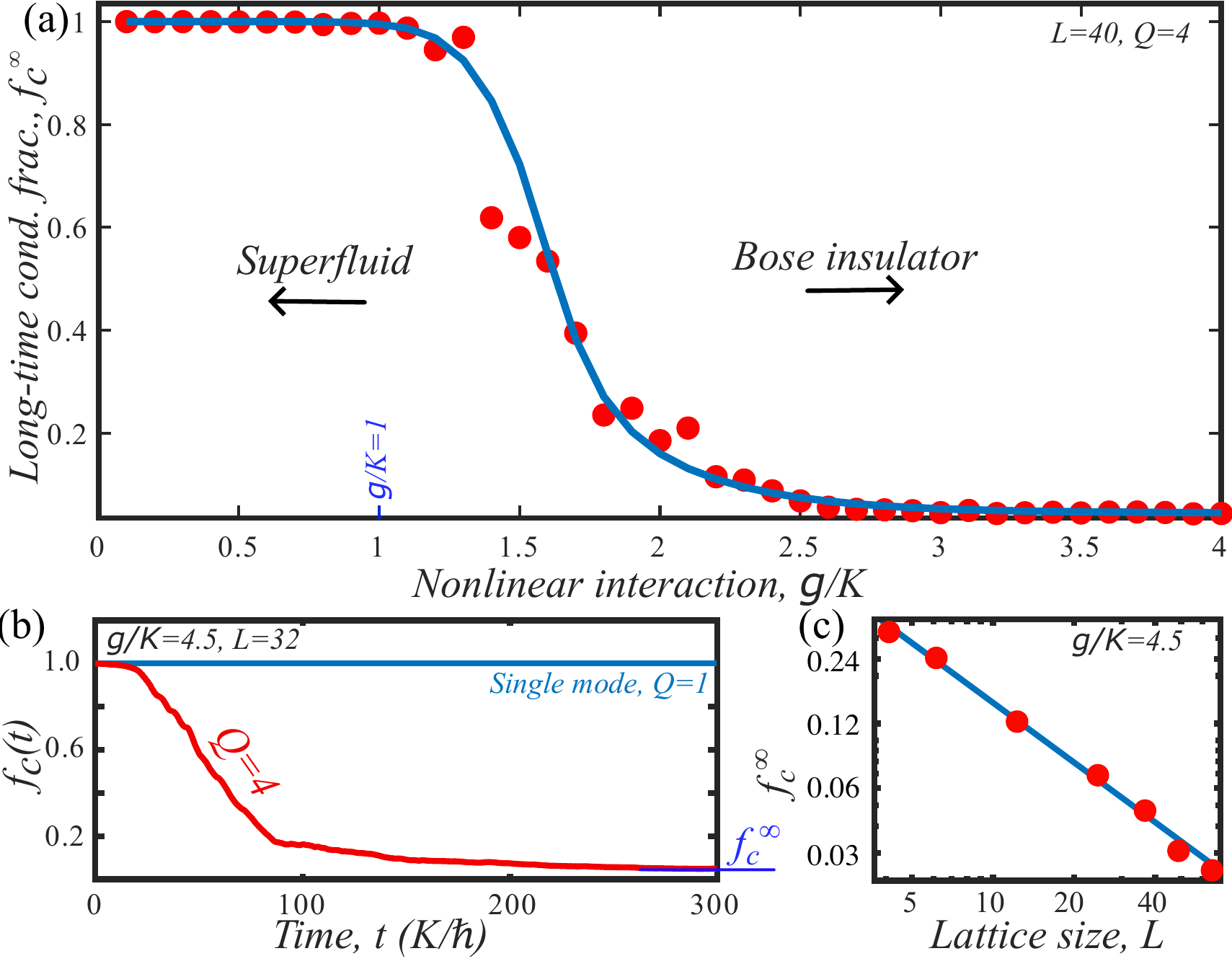}
\caption{Dynamical superfluid to Bose-insulator transition. (a) Long-time condensate fraction $f_c^\infty$ as a function of the interaction strength $g/K$. The condensate fraction suddenly drops at around $g/K\sim 1$. (b) The condensate fraction $f_c(t)$ as a function of time $t$ for a single-mode (blue, $Q=1$) and multi-mode (red, $Q=4$) configurations per site in the strong interaction regime $g/K = 4.5$. (c) Log-log plot of the long time condensate fraction $f_c^\infty$ as a function of the size $L$.}
\label{cond_fraction}
\end{figure}

\textbf{Long-Range Order:-- } Fig.~\ref{cond_fraction} shows the condensate fraction in a lattice with $L$ sites and $Q$ quantized levels per site. The \emph{condensate fraction} measures macroscopic coherence in interacting Bosonic systems~\cite{Penrose1956,Yang1962}, while \emph{off-diagonal long-range order} (ODLRO) characterizes Bose–Einstein condensation beyond the ideal-gas limit. Here, the central object of interest is the \emph{one-body density matrix}, defined as $\rho_{ij} = \langle \hat{\Psi}_i^\dagger \hat{\Psi}_j \rangle$ where $\hat{\Psi}_i$ is the Bosonic field operator at the lattice site $i$. The mean-field  time dependent field amplitudes $\psi_i(t)$ can be used to construct the corresponding one-body density matrix as $\rho_{ij}(t) = \langle \psi_i^*(t)\psi_j(t) \rangle_t$ from which a time-averaged density matrix is computed over a  temporal window $[0,t]$. 

The matrix $\rho_{ij}(t)$ is Hermitian and positive semi-definite, and can therefore be diagonalized as $\rho(t) = \sum_{a} n_{a}(t) |\zeta_{a}(t)\rangle \langle \zeta_{a} (t)|$, where $n_{a}$ are its eigenvalues (occupation numbers) and $|\zeta_{a}\rangle$ are the corresponding eigenmodes, often referred to as natural orbitals. The existence of a single macroscopic eigenvalue $n_0 = \mathcal{O}(N)$, scaling with the total number of particles $N$, signifies the emergence of ODLRO and the formation of a coherent condensate mode. The \emph{condensate fraction} is then defined as
\begin{equation}
f_c(t) = \frac{n_0(t)}{N},
\end{equation}
where $N = \mathrm{Tr}(\rho)$ denotes the total particle number~\cite{Fisher1989}. In the thermodynamic limit, a finite $f_c(t)$ indicates the presence of a condensate and thus superfluid order, while $\lim_{t\to\infty, L\to \infty} f_c(t) \to 0$ corresponds to an insulating phase in which long-range coherence is lost.

Fig.~\ref{cond_fraction}(a) shows a large condensate fraction ($\sim 1$) at weak nonlinear regime $g/K<1$, indicating a robust superfluid phase. However, a true superfluid phase with long-range order requires to overcome the long-wavelength phase fluctuations with a true symmetry breaking mechanism.

\textbf{U(1) gauge symmetry:-- } At weak interactions, the superfluid phase corresponds to a condensate occupying the zero quasi-momentum state at the bottom of the lattice band, characterized by global phase coherence and a finite superfluid stiffness ($\sim 2K f_c$). The corresponding Bogoliubov spectrum consists of linearly dispersing phonon modes, ensured by the positivity of both the compressibility and the effective mass at the band minimum~\cite{Utsunomiya2008}. In this regime, higher-energy states effectively introduce decay channels for the ground state. As a result, density fluctuations can be adiabatically eliminated and a phase-only theory ($\hbar \dot{\theta} \propto \nabla^2 \theta$) can be obtained, which suggests that fluctuations are smoothed out over time, stabilizing the superfluid phase. However, the corresponding kernel, proportional to $q^2$, leads to an infrared-divergent phase correlator $\langle |\theta(q)|^2 \rangle \sim 1/q^2$ in low-dimensional systems. This divergence reflects the presence of a gapless Goldstone mode rather than an instability: long-wavelength phase fluctuations are soft but do not grow exponentially in time. As a result, while infrared phase fluctuations may suppress true long-range order in low dimensions, they do not compromise the dynamical stability of the superfluid phase.

\textbf{Emergence of a dynamical Bose-insulator:--} At strong nonlinearity ($g/K \gg 1$), in addition to leakage into higher modes, complex interference processes arising from inter-modal mixing become significant. This introduces an additional effective fluctuating (stochastic) term in the phase equation. As shown in Fig.~\ref{cond_fraction}(a), the condensate fraction exhibits a sharp change as the interaction strength is varied. This behavior reflects a microscopic transformation of the system from a superfluid phase to a non-condensed phase in which long-range order is lost. The resulting state is a dynamical insulating phase, analogous to the pseudogap phases in superconductors, where global coherence vanishes even though local onsite condensate features can persist, each with a different phase~\cite{Dziarmaga2002,Schneider2012}. 

In Fig.~\ref{cond_fraction}(b), we observe that for a large interaction strength, $g/K = 4.5$, an initially prepared condensate gradually loses long-range coherence, as evidenced by the decay of the condensate fraction at long times. This behavior can be captured by an effective phase-only description,
\begin{eqnarray}
\hbar \dot{\theta}_i = \Gamma \partial_x^2 \theta_i + \eta_i(t),
\label{Eq:phase_only}
\end{eqnarray}
which is valid in the long-wavelength limit, where $\theta_i$ is the phase associated with the condensate at site $i$. The first term describes the restoring force associated with the superfluid phase stiffness, while the second term represents a rapidly fluctuating contribution arising from dynamical mode mixing. Notably, this fluctuating contribution is active only when small but finite average populations occupy excited levels. The validity of Eq.~\ref{Eq:phase_only} is restricted to wavelengths shorter than the system size $L$. As a result, a finite residual condensate fraction persists at long times in finite systems, as seen in Fig.~\ref{cond_fraction}(b). However, this residual fraction decreases with increasing system size, scaling approximately as $\sim 1/L$ [Fig.~\ref{cond_fraction}(c)], indicating that the system becomes insulating in the thermodynamic limit~\cite{Kinjo2023}.

\begin{figure}[h]
\includegraphics[width=.9\linewidth]{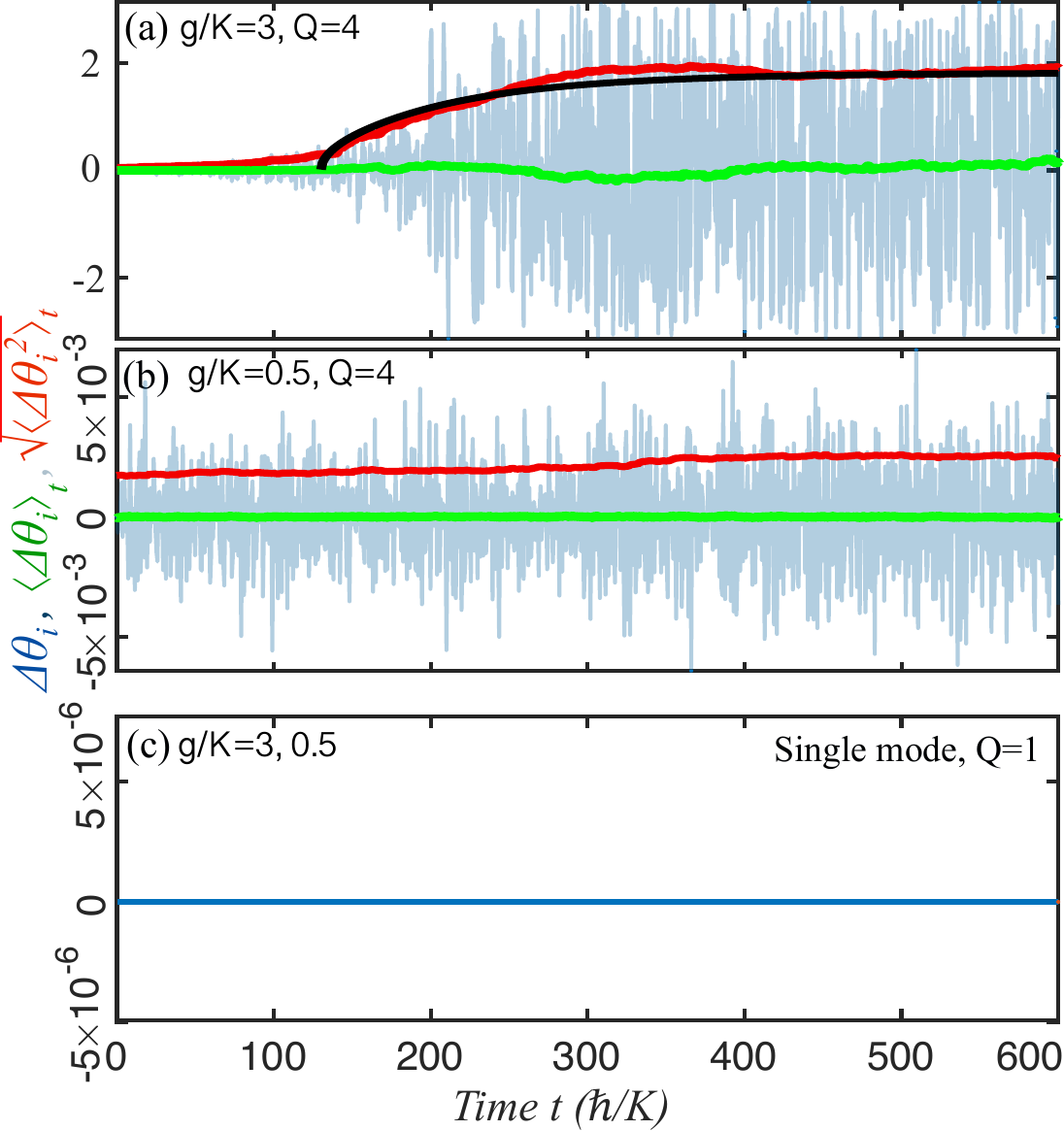}
\caption{Phase fluctuations induced by quantized levels. (a) Time dependence of the relative phase ($\Delta\theta_i$; light blue) time averaged phase ($\langle\Delta\theta_i\rangle_t$; green), and the second moment ($\langle\Delta\theta_i^2\rangle_t$; red) in the strong interaction regime $g/K=3$ and $Q=4$. The solid black line is the theoretical prediction for the second moment from Eq.~\ref{eq:second_moment}. (b) Time dependence of $\Delta\theta_i$ (light blue), $\langle\Delta\theta_i\rangle_t$ (green), and $\langle\Delta\theta_i^2\rangle_t$ (red) for weak interaction $g/K=0.5$ and $Q=4$. (c) For a single mode per site configuration $(Q=1)$, all $\Delta\theta_i$ (blue) remain zero for any interaction $g/K = 3$ or $0.5$.}
\label{Phase_diff}
\end{figure}

\textbf{Phase diffusion:-- } We find that the microscopic mechanism for the destruction of the global correlation observed in Fig.~\ref{cond_fraction} for strong nonlinear interactions is dominated by quantum fluctuations (see Fig.~\ref{Phase_diff})~\cite{Zinn2021}. Intriguingly, under this condition the relative phase $\Delta\theta_{i}$ turns out to be following the Brownian motion with the stochastic equation, $\hbar \Delta\dot{\theta}_{i}(t)\approx  \eta_i(t) +\omega^\text{eff}_i$, where $\eta_i(t)$ and $\omega^\text{eff}_i$ represent the rapidly fluctuating (due to mode mixing) and steady parts of the dynamical contributions respectively. The corresponding probability density $p(\Delta\theta_i,t)$ evolves according to the diffusion equation on a unit circle,
\begin{equation}
\partial_t p(\Delta\theta_i,t)=D\,\partial_{\Delta\theta_i}^2 p(\Delta\theta_i,t),
\end{equation}
with periodic boundary conditions $p(\Delta\theta_i,t)=p(\Delta\theta_i,+2\pi,t)$, where $D$ is the diffusion constant. Starting from an initial phase $\Delta\theta_i=0$, the solution can be expressed as a Fourier series, $2\pi p(\Delta\theta_i,t)=1+2\sum_{n=1}^\infty e^{-Dn^2 t}\cos(n\Delta\theta_i)$ related to the wrapped normal distributions~\cite{Mardia2009}. The variance is given by the second moment $\langle \Delta\theta_i^2\rangle$, which becomes
\begin{equation}
\langle \Delta\theta_i^2\rangle
=\frac{\pi^2}{3}+\frac{4}{\pi}\sum_{n=1}^{\infty}\frac{(-1)^n}{n^2}\,
e^{-Dn^2 (t-t_0)}
\label{eq:second_moment}
\end{equation}
for $t>t_0$, where $t_0$ is an initial time delay required to start the phase diffusion process. This expression shows explicitly how $\langle \Delta\theta_i^2\rangle$ grows linearly $\sim 2Dt$ at short times with an estimate $D\approx 0.01 K/\hbar$. Notably, $\sqrt{\langle \Delta\theta_{i}^2\rangle_t }$ approaches to a constant universal value $\pi/\sqrt{3}$ (see Fig.~\ref{Phase_diff}(a)), indicating complete randomization of the phase in the range $[-\pi,\pi]$ leading to destruction of the global coherence, and the restoration of  $U(1)$ gauge symmetry. For small interaction, phase fluctuation is limited to a small constant range $\sim \pm 10^{-3}$ (see Fig.~\ref{Phase_diff}(b)) leading to a strong global coherence (as shown in Fig.~\ref{cond_fraction}). 

Fig.~\ref{Phase_diff}(c) shows a constant dynamical phase with a diffusion constant $D=0$ for a single mode per site ($Q=1$), indicating no phase diffusion for any value of the nonlinearity $g$. This reinforces the importance of the multiple quantum levels $(Q>1)$ due to the Hilbert space quantization.

\textbf{Transitions vs crossover:--} In the weakly interacting regime, the system remains dynamically stable and supports a gapless Goldstone mode despite infrared-divergent phase fluctuations, owing to the stabilizing influence of higher modes. This situation changes once dynamical fluctuations are generated through mode mixing among multiple quantum levels. These dynamically generated fluctuations continuously inject phase noise and couple directly to the Goldstone mode, enhancing infrared fluctuations beyond what can be compensated by the phase-stiffness term. As a result, the two dynamically stable phases are qualitatively distinct and cannot be connected perturbatively. The observed behavior therefore corresponds to a dynamical phase transition rather than a smooth crossover.

We note, however, that in low-dimensional systems true long-range order may be absent due to infrared phase fluctuations associated with the gapless Goldstone mode. In a strict mathematical sense, this would suggest a sharp crossover rather than a phase transition characterized by a singularity. In higher dimensions, these conceptual limitations no longer apply.

\begin{figure} [h]
    \centering
    \includegraphics[width=0.9\linewidth]{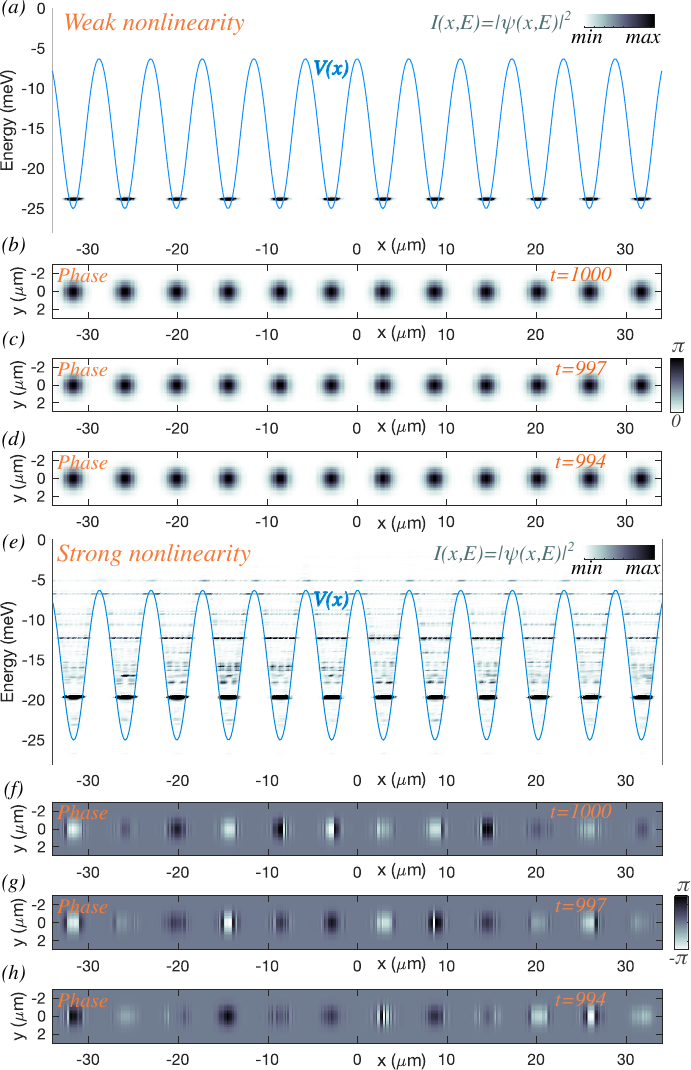}
    \caption{Continuous-model energy-resolved intensity.(a) The energy-resolved intensity exhibits a strong signal only at the lowest-energy state for weak nonlinearity ($P=0.3\, meV$). (b) The energy-resolved intensity shows significant contributions at multiple energy levels within each lattice site when the nonlinearity is large ($P=2.3\, meV$). We consider $\gamma = 0.2\, meV$, $g= (0.02 - i 0.01)\, meV/\mu m$, $V(x)= -V_0\sum_n \exp(-(x-nd_0)^2/\sigma^2)$, $V_0=25\, meV$, $d_0=5.76\,\mu m$, and $\sigma = 2\, \mu m$.}
    \label{ContinuousDispersions}
\end{figure}

\textbf{Continuous model.—} In Fig.~\ref{ContinuousDispersions} we examine whether the observed phenomena persist under realistic driven--dissipative conditions. In this framework, the effective nonlinearity is controlled by the particle number, set by the pumping strength \(P>\gamma\). Stabilization of the steady-state population is achieved by including a nonlinear loss term through the imaginary part of the interaction, \(g=g_R-i g_I\)~\cite{Keeling2008}.

For the fully driven--dissipative model, the wavefunction is initialized with a small amplitude and subsequently amplified by an incoherent pump \(P>\gamma\), driving the system to a finite condensate population where nonlinear effects become relevant. The steady-state population is stabilized by the nonlinear loss term \(\mathrm{Im}[g]\). Figures~\ref{ContinuousDispersions}(a,e) show the distribution of the local condensate population among quantized energy levels. In the weakly nonlinear regime (\(P=0.3~\mathrm{meV}\)), the population resides entirely in the ground state of each potential minimum, leading to suppressed phase fluctuations and coherent phase evolution across all sites. This is confirmed in Figs.~\ref{ContinuousDispersions}(b--d), where all local condensates exhibit identical phases at long times, up to an overall dynamical phase factor \(e^{-iE_0 t/\hbar}\).

The situation changes qualitatively in the strongly nonlinear regime. As shown in Fig.~\ref{ContinuousDispersions}(e), the local condensate population becomes broadly distributed among the quantized energy levels within each potential minimum. Consequently, multiple local modes actively participate in the dynamics, opening additional channels for phase fluctuations. While the presence of multiple quantized levels can stabilize the local phase at weak nonlinearity and support global phase coherence, strong nonlinearity instead enhances phase fluctuations, leading to the breakdown of global phase coherence, as evidenced in Figs.~\ref{ContinuousDispersions}(f--h).

\textbf{Discussions:--} Since the quantization of onsite energy plays the central role, all our discussions are restricted to the regime where the gaps between the levels $\Delta E \gg \gamma$. However, in the strongly driven-dissipative regime, new physics (e.g., the Kardar-Parisi-Zhang dynamics~\cite{Deligiannis2022,Novokreschenov2025}) may arise due to the relevance of the higher order corrections in the phase equation. From the practicality side, microcavity lattices with deep potential minima supporting multiple quantized energy levels have been demonstrated in several recent experiments~\cite{Amo2016,Su2020,Dusel2020,Gagel2024}. Moreover, condensates occupying multiple internal modes of local potentials have also been reported~\cite{Aladinskaia2023,Xiong2024,Wei2022,Zhang2019,Song2025}, with level spacings well above the polariton linewidth. These observations indicate that the mechanism we propose for a dynamical superfluid–Bose-insulator transition—driven by population of excited on-site modes and the resulting phase diffusion—can realistically be realized in near-term microcavity lattice systems.

\section{ACKNOWLEDGMENTS}
This work is supported by the National Natural Science Foundation of China (Grant No. 12274034), and the Quantum Science and Technology-National Science and Technology Major Project (2023ZD0300300).

\bibliography{main_arxiv}

@article{Sachdev1999,
  title={Quantum phase transitions},
  author={Sachdev, Subir},
  journal={Physics world},
  volume={12},
  number={4},
  pages={33},
  year={1999},
  publisher={IOP Publishing}
}

@article{Sondhi1997,
  title = {Continuous quantum phase transitions},
  author = {Sondhi, S. L. and Girvin, S. M. and Carini, J. P. and Shahar, D.},
  journal = {Rev. Mod. Phys.},
  volume = {69},
  issue = {1},
  pages = {315--333},
  numpages = {0},
  year = {1997},
  month = {Jan},
  publisher = {American Physical Society},
  doi = {10.1103/RevModPhys.69.315},
  url = {https://link.aps.org/doi/10.1103/RevModPhys.69.315}
}

@article{Vojta2003,
doi = {10.1088/0034-4885/66/12/R01},
url = {https://doi.org/10.1088/0034-4885/66/12/R01},
year = {2003},
month = {nov},
publisher = {},
volume = {66},
number = {12},
pages = {2069},
author = {Matthias Vojta},
title = {Quantum phase transitions},
journal = {Reports on Progress in Physics},
}

@article{Wang_2024,
doi = {10.1088/1361-6633/ad14f3},
url = {https://doi.org/10.1088/1361-6633/ad14f3},
year = {2023},
month = {dec},
publisher = {IOP Publishing},
volume = {87},
number = {1},
pages = {014502},
author = {Wang, Ziqiao and Liu, Yi and Ji, Chengcheng and Wang, Jian},
title = {Quantum phase transitions in two-dimensional superconductors: a review on recent experimental progress},
journal = {Reports on Progress in Physics}
}

@article{Bloch2022,
  title={The superfluid-to-Mott insulator transition and the birth of experimental quantum simulation},
  author={Bloch, Immanuel and Greiner, Markus},
  journal={Nature Reviews Physics},
  volume={4},
  number={12},
  pages={739--740},
  year={2022},
  publisher={Nature Publishing Group UK London}
}

@article{Penrose1956,
  title = {Bose-Einstein Condensation and Liquid Helium},
  author = {Penrose, Oliver and Onsager, Lars},
  journal = {Phys. Rev.},
  volume = {104},
  issue = {3},
  pages = {576--584},
  numpages = {0},
  year = {1956},
  month = {Nov},
  publisher = {American Physical Society},
  doi = {10.1103/PhysRev.104.576},
  url = {https://link.aps.org/doi/10.1103/PhysRev.104.576}
}

@article{Yang1962,
  title = {Concept of Off-Diagonal Long-Range Order and the Quantum Phases of Liquid He and of Superconductors},
  author = {Yang, C. N.},
  journal = {Rev. Mod. Phys.},
  volume = {34},
  issue = {4},
  pages = {694--704},
  numpages = {0},
  year = {1962},
  month = {Oct},
  publisher = {American Physical Society},
  doi = {10.1103/RevModPhys.34.694},
  url = {https://link.aps.org/doi/10.1103/RevModPhys.34.694}
}

@article{Schneider2012,
  title = {Superconductor-Insulator Quantum Phase Transition in Disordered FeSe Thin Films},
  author = {Schneider, R. and Zaitsev, A. G. and Fuchs, D. and v. Lohneysen, H.},
  journal = {Phys. Rev. Lett.},
  volume = {108},
  issue = {25},
  pages = {257003},
  numpages = {5},
  year = {2012},
  month = {Jun},
  publisher = {American Physical Society},
  doi = {10.1103/PhysRevLett.108.257003},
  url = {https://link.aps.org/doi/10.1103/PhysRevLett.108.257003}
}

@article{Dziarmaga2002,
  title = {Dynamics of Quantum Phase Transition in an Array of Josephson Junctions},
  author = {Dziarmaga, J. and Smerzi, A. and Zurek, W. H. and Bishop, A. R.},
  journal = {Phys. Rev. Lett.},
  volume = {88},
  issue = {16},
  pages = {167001},
  numpages = {4},
  year = {2002},
  month = {Apr},
  publisher = {American Physical Society},
  doi = {10.1103/PhysRevLett.88.167001},
  url = {https://link.aps.org/doi/10.1103/PhysRevLett.88.167001}
}

@article{Carusotto2013,
  title = {Quantum fluids of light},
  author = {Carusotto, Iacopo and Ciuti, Cristiano},
  journal = {Rev. Mod. Phys.},
  volume = {85},
  issue = {1},
  pages = {299--366},
  numpages = {0},
  year = {2013},
  month = {Feb},
  publisher = {American Physical Society},
  doi = {10.1103/RevModPhys.85.299},
  url = {https://link.aps.org/doi/10.1103/RevModPhys.85.299}
}

@article{Ghosh2022,
  title={Microcavity exciton polaritons at room temperature},
  author={Ghosh, Sanjib and Su, Rui and Zhao, Jiaxin and Fieramosca, Antonio and Wu, Jinqi and Li, Tengfei and Zhang, Qing and Li, Feng and Chen, Zhanghai and Liew, Timothy and others},
  journal={Photonics Insights},
  volume={1},
  number={1},
  pages={R04--R04},
  year={2022},
  publisher={Society of Photo-Optical Instrumentation Engineers}
}

@article{Greentree2006,
  title={Quantum phase transitions of light},
  author={Greentree, Andrew D and Tahan, Charles and Cole, Jared H and Hollenberg, Lloyd CL},
  journal={Nature Physics},
  volume={2},
  number={12},
  pages={856--861},
  year={2006},
  publisher={Nature Publishing Group UK London}
}

@article{Deng2010,
  title = {Exciton-polariton Bose-Einstein condensation},
  author = {Deng, Hui and Haug, Hartmut and Yamamoto, Yoshihisa},
  journal = {Rev. Mod. Phys.},
  volume = {82},
  issue = {2},
  pages = {1489--1537},
  numpages = {0},
  year = {2010},
  month = {May},
  publisher = {American Physical Society},
  doi = {10.1103/RevModPhys.82.1489},
  url = {https://link.aps.org/doi/10.1103/RevModPhys.82.1489}
}

@article{Byrnes2014,
  title={Exciton--polariton condensates},
  author={Byrnes, Tim and Kim, Na Young and Yamamoto, Yoshihisa},
  journal={Nature Physics},
  volume={10},
  number={11},
  pages={803--813},
  year={2014},
  publisher={Nature Publishing Group UK London}}

@article{Brune2025,
  title={Quantum coherence of a long-lifetime exciton-polariton condensate},
  author={Brune, Yannik and Rozas, Elena and West, Ken and Baldwin, Kirk and Pfeiffer, Loren N and Beaumariage, Jonathan and Alnatah, Hassan and Snoke, David W and A{\ss}mann, Marc},
  journal={Communications Materials},
  volume={6},
  number={1},
  pages={123},
  year={2025},
  publisher={Nature Publishing Group UK London}
}

@article{Liew2023,
author = {T. C. H. Liew},
journal = {Opt. Mater. Express},
number = {7},
pages = {1938--1946},
publisher = {Optica Publishing Group},
title = {The future of quantum in polariton systems: opinion},
volume = {13},
month = {Jul},
year = {2023},
url = {https://opg.optica.org/ome/abstract.cfm?URI=ome-13-7-1938},
doi = {10.1364/OME.492503}
}

@article{Hu2024,
author = {Y.-M. Robin Hu and Elena A. Ostrovskaya and Eliezer Estrecho},
journal = {Opt. Mater. Express},
number = {3},
pages = {664--686},
publisher = {Optica Publishing Group},
title = {Generalized quantum geometric tensor in a non-Hermitian exciton-polariton system [Invited]},
volume = {14},
month = {Mar},
year = {2024},
url = {https://opg.optica.org/ome/abstract.cfm?URI=ome-14-3-664},
doi = {10.1364/OME.497010}
}

@article{Klembt2018,
  title={Exciton-polariton topological insulator},
  author={Klembt, Sebastian and Harder, TH and Egorov, OA and Winkler, K and Ge, R and Bandres, MA and Emmerling, M and Worschech, L and Liew, TCH and Segev, M and others},
  journal={Nature},
  volume={562},
  number={7728},
  pages={552--556},
  year={2018},
  publisher={Nature Publishing Group UK London}
}

@article{Jia2025,
    author = {Jia, Haoyuan and Cao, Junhui and Chen, Fei and Peng, Fangying and Li, Yihui and Xu, Yihan and Chen, Leizhu and Ye, Ziyu and Zhao, Xianyan and Zhang, Shian and Jing, Jietai and Xu, Hongxing and Chen, Zhanghai and Byrnes, Tim and Li, Hui and Kavokin, Alexey and Wu, Jian},
    title = {Femtosecond coherence dynamics of exciton polaritons},
    journal = {National Science Review},
    pages = {nwaf493},
    year = {2025},
    month = {11},
    issn = {2095-5138},
    doi = {10.1093/nsr/nwaf493},
    url = {https://doi.org/10.1093/nsr/nwaf493},
    eprint = {https://academic.oup.com/nsr/advance-article-pdf/doi/10.1093/nsr/nwaf493/65388832/nwaf493.pdf},
}

@article{Su2020,
  title={Observation of exciton polariton condensation in a perovskite lattice at room temperature},
  author={Su, Rui and Ghosh, Sanjib and Wang, Jun and Liu, Sheng and Diederichs, Carole and Liew, Timothy CH and Xiong, Qihua},
  journal={Nature Physics},
  volume={16},
  number={3},
  pages={301--306},
  year={2020},
  publisher={Nature Publishing Group UK London}
}

@article{Jean2017,
  title={Lasing in topological edge states of a one-dimensional lattice},
  author={St-Jean, Philippe and Goblot, Victor and Galopin, Elisabeth and Lema{\^\i}tre, A and Ozawa, Thomas and Le Gratiet, Luc and Sagnes, Isabelle and Bloch, Jacqueline and Amo, Alberto},
  journal={Nature Photonics},
  volume={11},
  number={10},
  pages={651--656},
  year={2017},
  publisher={Nature Publishing Group UK London}
}

@article{Amo2016,
title = {Exciton-polaritons in lattices: A non-linear photonic simulator},
journal = {Comptes Rendus Physique},
volume = {17},
number = {8},
pages = {934-945},
year = {2016},
note = {Polariton physics / Physique des polaritons},
issn = {1631-0705},
doi = {https://doi.org/10.1016/j.crhy.2016.08.007},
url = {https://www.sciencedirect.com/science/article/pii/S163107051630086X},
author = {Alberto Amo and Jacqueline Bloch}}

@article{Wu2023,
  title={Higher-order topological polariton corner state lasing},
  author={Wu, Jinqi and Ghosh, Sanjib and Gan, Yusong and Shi, Ying and Mandal, Subhaskar and Sun, Handong and Zhang, Baile and Liew, Timothy CH and Su, Rui and Xiong, Qihua},
  journal={Science Advances},
  volume={9},
  number={21},
  pages={eadg4322},
  year={2023},
  publisher={American Association for the Advancement of Science}
}

@article{Wouters2007,
  title = {Excitations in a Nonequilibrium Bose-Einstein Condensate of Exciton Polaritons},
  author = {Wouters, Michiel and Carusotto, Iacopo},
  journal = {Phys. Rev. Lett.},
  volume = {99},
  issue = {14},
  pages = {140402},
  numpages = {4},
  year = {2007},
  month = {Oct},
  publisher = {American Physical Society},
  doi = {10.1103/PhysRevLett.99.140402},
  url = {https://link.aps.org/doi/10.1103/PhysRevLett.99.140402}
}

@article{Keeling2008,
  title = {Spontaneous Rotating Vortex Lattices in a Pumped Decaying Condensate},
  author = {Keeling, Jonathan and Berloff, Natalia G.},
  journal = {Phys. Rev. Lett.},
  volume = {100},
  issue = {25},
  pages = {250401},
  numpages = {4},
  year = {2008},
  month = {Jun},
  publisher = {American Physical Society},
  doi = {10.1103/PhysRevLett.100.250401},
  url = {https://link.aps.org/doi/10.1103/PhysRevLett.100.250401}
}

@book{Mardia2009,
  title={Directional statistics},
  author={Mardia, Kanti V and Jupp, Peter E},
  year={2009},
  publisher={John Wiley \& Sons}
}

@article{Dusel2020,
  title={Room temperature organic exciton--polariton condensate in a lattice},
  author={Dusel, Marco and Betzold, Simon and Egorov, Oleg A and Klembt, Sebastian and Ohmer, J{u}rgen and Fischer, Utz and H{o}fling, Sven and Schneider, Christian},
  journal={Nature communications},
  volume={11},
  number={1},
  pages={2863},
  year={2020},
  publisher={Nature Publishing Group UK London}
}

@article{Aladinskaia2023,
  title = {Spatial quantization of exciton-polariton condensates in optically induced traps},
  author = {Aladinskaia, Ekaterina and Cherbunin, Roman and Sedov, Evgeny and Liubomirov, Alexey and Kavokin, Kirill and Khramtsov, Evgeny and Petrov, Mikhail and Savvidis, P. G. and Kavokin, Alexey},
  journal = {Phys. Rev. B},
  volume = {107},
  issue = {4},
  pages = {045302},
  numpages = {7},
  year = {2023},
  month = {Jan},
  publisher = {American Physical Society},
  doi = {10.1103/PhysRevB.107.045302},
  url = {https://link.aps.org/doi/10.1103/PhysRevB.107.045302}
}

@article{Xiong2024,
  title={Selective excitation of exciton--polariton condensate modes in an annular perovskite microcavity},
  author={Xiong, Zhenyu and Wu, Hao and Cai, Yuanwen and Zhai, Xiaokun and Liu, Tong and Li, Baili and Song, Tieling and Guo, Longfei and Liu, Zhengliang and Dong, Yifan and others},
  journal={Nano Letters},
  volume={24},
  number={16},
  pages={4959--4964},
  year={2024},
  publisher={ACS Publications}
}

@article{Wei2022,
  title={Optically trapped room temperature polariton condensate in an organic semiconductor},
  author={Wei, Mengjie and Verstraelen, Wouter and Orfanakis, Konstantinos and Ruseckas, Arvydas and Liew, Timothy CH and Samuel, Ifor DW and Turnbull, Graham A and Ohadi, Hamid},
  journal={Nature communications},
  volume={13},
  number={1},
  pages={7191},
  year={2022},
  publisher={Nature Publishing Group UK London}
}

@article{Gagel2024,
  title={An electrically pumped topological polariton laser},
  author={Gagel, Philipp and Egorov, Oleg A and Dzimira, Franciszek and Beierlein, Johannes and Emmerling, Monika and Wolf, Adriana and Jabeen, Fauzia and Betzold, Simon and Peschel, Ulf and Hofling, Sven and others},
  journal={Nano Letters},
  volume={24},
  number={22},
  pages={6538--6544},
  year={2024},
  publisher={ACS Publications}
}

@article{Zhang2019,
  title={Room temperature exciton--polariton condensate in an optically-controlled trap},
  author={Zhang, Xinhan and Zhang, Yingjun and Dong, Hongxing and Tang, Bing and Li, Dehui and Tian, Chuan and Xu, Chunyan and Zhou, Weihang},
  journal={Nanoscale},
  volume={11},
  number={10},
  pages={4496--4502},
  year={2019},
  publisher={Royal Society of Chemistry}
}

@article{Song2025,
  title={Room-temperature continuous-wave pumped exciton polariton condensation in a perovskite microcavity},
  author={Song, Jiepeng and Ghosh, Sanjib and Deng, Xinyi and Li, Chun and Shang, Qiuyu and Liu, Xinfeng and Wang, Yubin and Gao, Xiaoyue and Yang, Wenkai and Wang, Xianjin and others},
  journal={Science Advances},
  volume={11},
  number={5},
  pages={eadr1652},
  year={2025},
  publisher={American Association for the Advancement of Science}
}

@article{Byrnes2010,
  title = {Mott transitions of exciton polaritons and indirect excitons in a periodic potential},
  author = {Byrnes, Tim and Recher, Patrik and Yamamoto, Yoshihisa},
  journal = {Phys. Rev. B},
  volume = {81},
  issue = {20},
  pages = {205312},
  numpages = {13},
  year = {2010},
  month = {May},
  publisher = {American Physical Society},
  doi = {10.1103/PhysRevB.81.205312},
  url = {https://link.aps.org/doi/10.1103/PhysRevB.81.205312}
}

@article{Na2010,
  title={Massive parallel generation of indistinguishable single photons via the polaritonic superfluid to Mott-insulator quantum phase transition},
  author={Na, Neil and Yamamoto, Yoshihisa},
  journal={New Journal of Physics},
  volume={12},
  number={12},
  pages={123001},
  year={2010},
  publisher={IOP Publishing}
}

@article{Krol2025,
  title={Momentum-space non-Hermitian skin effect in an exciton-polariton system},
  author={Kr{\'o}l, Mateusz and Smirnova, Daria A and Smirnov, Lev A and Fabricante, Bianca Rae and Winkler, Karol and Kamp, Martin and Schneider, Christian and H{o}fling, Sven and Liew, Timothy CH and Truscott, Andrew G and others},
  journal={arXiv preprint arXiv:2512.10146},
  year={2025}
}

@article{Bao2025,
  title={Exciton-Polariton hybrid skin-topological states},
  author={Bao, Ruiqi and Banerjee, R and Mandal, S and Xu, Huawen and Li, Shiji and Gao, Junfeng and Liew, Timothy CH},
  journal={arXiv preprint arXiv:2512.01768},
  year={2025}
}

@article{Xu2025,
  title = {Exciton polariton critical non-Hermitian skin effect with spin-momentum-locked gains},
  author = {Xu, Xingran and Tian, Lingyu and An, Zhiyuan and Xiong, Qihua and Ghosh, Sanjib},
  journal = {Phys. Rev. B},
  volume = {111},
  issue = {12},
  pages = {L121301},
  numpages = {7},
  year = {2025},
  month = {Mar},
  publisher = {American Physical Society},
  doi = {10.1103/PhysRevB.111.L121301},
  url = {https://link.aps.org/doi/10.1103/PhysRevB.111.L121301}
}

@article{Bardyn2015,
  title = {Topological polaritons and excitons in garden-variety systems},
  author = {Bardyn, Charles-Edouard and Karzig, Torsten and Refael, Gil and Liew, Timothy C. H.},
  journal = {Phys. Rev. B},
  volume = {91},
  issue = {16},
  pages = {161413},
  numpages = {5},
  year = {2015},
  month = {Apr},
  publisher = {American Physical Society},
  doi = {10.1103/PhysRevB.91.161413},
  url = {https://link.aps.org/doi/10.1103/PhysRevB.91.161413}
}

@article{Karzig2015,
  title = {Topological Polaritons},
  author = {Karzig, Torsten and Bardyn, Charles-Edouard and Lindner, Netanel H. and Refael, Gil},
  journal = {Phys. Rev. X},
  volume = {5},
  issue = {3},
  pages = {031001},
  numpages = {10},
  year = {2015},
  month = {Jul},
  publisher = {American Physical Society},
  doi = {10.1103/PhysRevX.5.031001},
  url = {https://link.aps.org/doi/10.1103/PhysRevX.5.031001}
}

@book{Kavokin2017,
  title={Microcavities},
  author={Kavokin, Alexey and Baumberg, Jeremy J and Malpuech, Guillaume and Laussy, Fabrice P},
  year={2017},
  publisher={Oxford university press}
}

@article{Masharin2025,
  title={Non-Hermitian trapping of Dirac exciton-polariton condensates in a perovskite metasurface},
  author={Masharin, Mikhail and Chestnov, Igor and Bochin, Andrey and Kozhevin, Pavel and Shahnazaryan, Vanik and Yulin, Alexey and Iorsh, Ivan and Ma, Xuekai and Schumacher, Stefan and Makarov, Sergey and others},
  journal={arXiv preprint arXiv:2512.09603},
  year={2025}
}

@article{SongKok2025,
  title = {Electrically Tunable and Enhanced Nonlinearity of Moir\'e Exciton Polaritons in Transition Metal Dichalcogenide Bilayers},
  author = {Song, Kok Wee and Kyriienko, Oleksandr},
  journal = {Phys. Rev. Lett.},
  volume = {135},
  issue = {3},
  pages = {036901},
  numpages = {8},
  year = {2025},
  month = {Jul},
  publisher = {American Physical Society},
  doi = {10.1103/gr72-szwg},
  url = {https://link.aps.org/doi/10.1103/gr72-szwg}
}

@article{Matuszewski2024,
  title = {Role of all-optical neural networks},
  author = {Matuszewski, M. and Prystupiuk, A. and Opala, A.},
  journal = {Phys. Rev. Appl.},
  volume = {21},
  issue = {1},
  pages = {014028},
  numpages = {15},
  year = {2024},
  month = {Jan},
  publisher = {American Physical Society},
  doi = {10.1103/PhysRevApplied.21.014028},
  url = {https://link.aps.org/doi/10.1103/PhysRevApplied.21.014028}
}

@article{Opala2025,
  title={Perovskite Microwires for Room Temperature Exciton-Polariton Neural Network},
  author={Opala, Andrzej and Tyszka, Krzysztof and K{{e}}dziora, Mateusz and Furman, Magdalena and Rahmani, Amir and {\'S}wierczewski, Stanis{\l}aw and Ekielski, Marek and Szerling, Anna and Matuszewski, Micha{\l} and Pi{{e}}tka, Barbara},
  journal={Advanced Materials},
  volume={37},
  number={43},
  pages={e07612},
  year={2025},
  publisher={Wiley Online Library}
}

@article{Wouter2025,
  title = {Optical resonators constitute a universal spin simulator},
  author = {Verstraelen, Wouter and Liew, Timothy C. H.},
  journal = {Phys. Rev. B},
  volume = {111},
  issue = {1},
  pages = {014307},
  numpages = {7},
  year = {2025},
  month = {Jan},
  publisher = {American Physical Society},
  doi = {10.1103/PhysRevB.111.014307},
  url = {https://link.aps.org/doi/10.1103/PhysRevB.111.014307}
}

@article{Shi2025,
  title = {Chiral Dissociation of Bound Photon Pairs for a Non-Hermitian Skin Effect},
  author = {Shi, Jiaming and Poddubny, Alexander N.},
  journal = {Phys. Rev. Lett.},
  volume = {134},
  issue = {23},
  pages = {233602},
  numpages = {6},
  year = {2025},
  month = {Jun},
  publisher = {American Physical Society},
  doi = {10.1103/q6wr-2rt9},
  url = {https://link.aps.org/doi/10.1103/q6wr-2rt9}
}

@article{Barrat2024,
  title={Qubit analog with polariton superfluid in an annular trap},
  author={Barrat, Joris and Tzortzakakis, Andreas F and Niu, Meng and Zhou, Xiaoqing and Paschos, Giannis G and Petrosyan, David and Savvidis, Pavlos G},
  journal={Science Advances},
  volume={10},
  number={43},
  pages={eado4042},
  year={2024},
  publisher={American Association for the Advancement of Science}
}

@article{Fisher1989,
  title = {Boson localization and the superfluid-insulator transition},
  author = {Fisher, Matthew P. A. and Weichman, Peter B. and Grinstein, G. and Fisher, Daniel S.},
  journal = {Phys. Rev. B},
  volume = {40},
  issue = {1},
  pages = {546--570},
  numpages = {0},
  year = {1989},
  month = {Jul},
  publisher = {American Physical Society},
  doi = {10.1103/PhysRevB.40.546},
  url = {https://link.aps.org/doi/10.1103/PhysRevB.40.546}
}

@book{Zinn2021,
  title={Quantum field theory and critical phenomena},
  author={Zinn-Justin, Jean},
  volume={171},
  year={2021},
  publisher={Oxford university press}
}

@article{Kinjo2023,
  title={Dynamics of quantum double dark-solitons and an exact finite-size scaling of Bose--Einstein condensation},
  author={Kinjo, Kayo and Sato, Jun and Deguchi, Tetsuo},
  journal={Journal of Physics A: Mathematical and Theoretical},
  volume={56},
  number={16},
  pages={164001},
  year={2023},
  publisher={IOP Publishing}
}

@article{Sedov2025,
  title={Polariton lattices as binarized neuromorphic networks},
  author={Sedov, Evgeny and Kavokin, Alexey},
  journal={Light: Science \& Applications},
  volume={14},
  number={1},
  pages={52},
  year={2025},
  publisher={Nature Publishing Group UK London}
}

@article{Dagvadorj2015,
  title = {Nonequilibrium Phase Transition in a Two-Dimensional Driven Open Quantum System},
  author = {Dagvadorj, G. and Fellows, J. M. and Matyja\ifmmode \acute{s}\else \'{s}\fi{}kiewicz, S. and Marchetti, F. M. and Carusotto, I. and Szyma\ifmmode \acute{n}\else \'{n}\fi{}ska, M. H.},
  journal = {Phys. Rev. X},
  volume = {5},
  issue = {4},
  pages = {041028},
  numpages = {9},
  year = {2015},
  month = {Nov},
  publisher = {American Physical Society},
  doi = {10.1103/PhysRevX.5.041028},
  url = {https://link.aps.org/doi/10.1103/PhysRevX.5.041028}
}

@book{Domb2000,
  title={Phase transitions and critical phenomena},
  author={Domb, Cyril},
  volume={19},
  year={2000},
  publisher={Elsevier}
}

@article{Utsunomiya2008,
  title={Observation of Bogoliubov excitations in exciton-polariton condensates},
  author={Utsunomiya, S and Tian, L and Roumpos, G and Lai, CW and Kumada, N and Fujisawa, T and Kuwata-Gonokami, M and L{o}ffler, A and H{o}fling, Sven and Forchel, A and others},
  journal={Nature Physics},
  volume={4},
  number={9},
  pages={700--705},
  year={2008},
  publisher={Nature Publishing Group UK London}
}

@article{Deligiannis2022,
  title = {Kardar-Parisi-Zhang universality in discrete two-dimensional driven-dissipative exciton polariton condensates},
  author = {Deligiannis, Konstantinos and Fontaine, Quentin and Squizzato, Davide and Richard, Maxime and Ravets, Sylvain and Bloch, Jacqueline and Minguzzi, Anna and Canet, L\'eonie},
  journal = {Phys. Rev. Res.},
  volume = {4},
  issue = {4},
  pages = {043207},
  numpages = {13},
  year = {2022},
  month = {Dec},
  publisher = {American Physical Society},
  doi = {10.1103/PhysRevResearch.4.043207},
  url = {https://link.aps.org/doi/10.1103/PhysRevResearch.4.043207}
}

@article{Novokreschenov2025,
  title = {Kardar-Parisi-Zhang universality in optically induced lattices of exciton-polariton condensates},
  author = {Novokreschenov, D. and Neplokh, V. and Misko, M. and Starkova, N. and Cookson, T. and Kudlis, A. and Nalitov, A. and Shelykh, I. A. and Kavokin, A. V. and Lagoudakis, P.},
  journal = {Phys. Rev. B},
  volume = {112},
  issue = {17},
  pages = {174313},
  numpages = {8},
  year = {2025},
  month = {Nov},
  publisher = {American Physical Society},
  doi = {10.1103/ym2g-t8vj},
  url = {https://link.aps.org/doi/10.1103/ym2g-t8vj}
}

\end{document}